# Near-infrared metalens empowered dual-mode high resolution and large FOV microscope

Chuang Sun[1], Hailong Pi[1], Kian Shen Kiang[1], Jize Yan[1,]*, and Jun-Yu Ou[2,]*

[1]School of Electronics and Computer Science, University of Southampton, Southampton SO17 1BJ, UK
[2]Department of Physics and Astronomy, University of Southampton, Southampton SO17 1BJ, UK

*J.Yan@soton.ac.uk ; bruce.ou@soton.ac.uk

**Abstract:** The spiral phase contrast microscope can clearly distinguish the morphological information of the low contrast objects (i.e., biological samples) because of the isotropic edge-enhancement effect, while the bright field microscope can image the overall morphology of amplitude objects. However, the imaging resolution, magnification, and field of view of conventional spiral phase contrast microscopes based on 4f filtering configuration are limited by the system's complexity. Here, we reported compact dual-mode microscopes working at near-infrared using the engineered metalens which can be tuned between the spiral phase contrast imaging and bright field imaging by polarization control. The metalens combines the high-resolution objective lens and polarization-controlled phase filter into a single-layer nanofins array. We demonstrated two infinity-corrected microscope systems to achieve subwavelength resolution ($0.7\lambda$), large magnification (58X), and large field of view (600μm× 800μm). Unstained onion epidermal is imaged by the microscope to show the dual-mode imaging ability for the biological sample. Finally, a singlet dual-mode microscope system is demonstrated to show the edge-detection application for industrial standards. Our results could open new opportunities in applications of biological imaging, industrial machine vision, and semiconductor inspection..

## 1. Introduction

Bright field microscope is widely used for imaging the overall morphologies of amplitude objects. However, the bright field microscope cannot provide clearly distinguished morphological information of the transparent objects (i.e., low index contrast objects) [1]. To realize high-contrast and edge-enhanced imaging of phase objects, a phase contrast microscope is initially introduced by the Dutch physicist Frits Zernike in the 1930s [2, 3]. Compared with algorithm-based edge enhancement in image processing, a phase contrast microscope is an all-optical processing and could provide direct edge detection functionality [4]. This technique is employed to generate high-contrast images of transparent specimens, including living cells (typically in culture), unstained biological samples, and microorganisms and is widely used for low optical contrast specimens [5].

Essentially, a phase contrast microscope is based on filtering image information in its Fourier plane (i.e., frequency domain) [3]. As the spiral phase filtering function can lead to a 2-dimensional (2D) isotropic edge-enhancement effect of observed amplitude and phase objects, the spiral phase contrast microscope has been implemented in many ways. Conventionally, the spiral phase contrast microscope is implemented by generating a spiral phase profile in the Fourier plane of a 4f filtering system via a spatial light modulate (SLM), which makes the microscope system too bulky and limits the resolution and field of view (FOV) [6, 7].

In recent years, metasurface being made of an array of subwavelength nanofins has attracted a lot of attention in the imaging community because of its powerful ability to manipulate amplitude, phase, and polarization of incident light [8-11]. A transmissive dielectric metasurface working as a polarization-controlled phase filter is placed in the Fourier plane of a 4f system for realizing switchable spiral phase contrast imaging [12]. Additionally, some reflection configurations based on reflective metasurface are proposed as well [13-14]. However, the multiple lenses used in both 4f filtering system or reflection configuration result in a complex and bulky optical system, which fails to exploit the multi-functional merit of a metasurface and hinders the miniaturization of the spiral phase contrast imaging system.

A single-lens spiral phase contrast imaging configuration was achieved by compressing the imaging and edge-enhancement functionalities into transmissive metalens in the visible spectrum [15]. However, the metalens can only work in the phase-contrast imaging mode, which limits its application in bright field imaging mode. Recently, a dielectric metasurface has been proposed for synchronous spiral phase contrast and bright field imaging, and an electrically tunable dual-mode metalens is reported for switchable spiral phase contrast and bright field imaging [16]. However, the former one has one small imaging FOV (~$50\mu m \times 50\mu m$), and the latter one has a low resolution (~$3\lambda$) and low edge-enhancement quality. There is no report on the switchable dual-mode (i.e., bright field and phase contrast) microscope with high resolution, large magnification, and large FOV. While the above metalenses work in the visible spectrum., the near-infrared (NIR) spectrum is attracting increasing attention in biological research, machine vision and semiconductor inspection because of several advantages. Firstly, the NIR light wave can penetrate through the biological tissue deeper than visible light. Secondly, the visible auto-fluorescence of cells and tissues is considered as the main source of background noise in biological imaging, imaging in NIR can achieve a higher signal-to-noise ratio [17, 18]. Finally, the NIR microscopy provides defect information on the silicon chips compared to visible light for semiconductor inspection.

Here, we propose a polarization-controlled single-layer dielectric metalens where the phase profile can be tuned from a hyperbolic phase to a sum of a hyperbolic phase and a spiral phase with a topologic charge of 1, as shown in Figure 1. In the wavelength of 1550nm, the metalens can focus the left-handed circular polarized(LCP) light beam to a donut ring with a topologic charge of 1 for phase contrast imaging mode and focus the right-handed circular polarized(RCP) light beam to a Gaussian focal point for bright field imaging. Taking these advantages, we demonstrate polarization-controlled dual-mode imaging systems in the NIR imaging window. The design and working principle of the metalens are derived at first, and two metalens samples with a numerical aperture (NA) of 0.89 and 0.25 are fabricated and characterized in the experiment.

## 2. Metalens design and characterization

We use the spin-multiplexing method imposing two distinct phase profiles on the LCP and RCP light beams in the same metalens, respectively. As shown in Fig. 1(a), to realize phase contrast imaging mode, the metalens impart a sum phase profile [Eq. (1)] of the hyperbolic phase and the spiral phase with a topological charge of 1 on the LCP light beam. A hyperbolic phase profile [Eq. (2)] is imparted onto the RCP light beam to realize bright field imaging mode. Therefore, the metalens in a microscope system can serve as an imaging lens and a polarization-controlled phase plate as shown in Fig. 1(b) and 1(c).

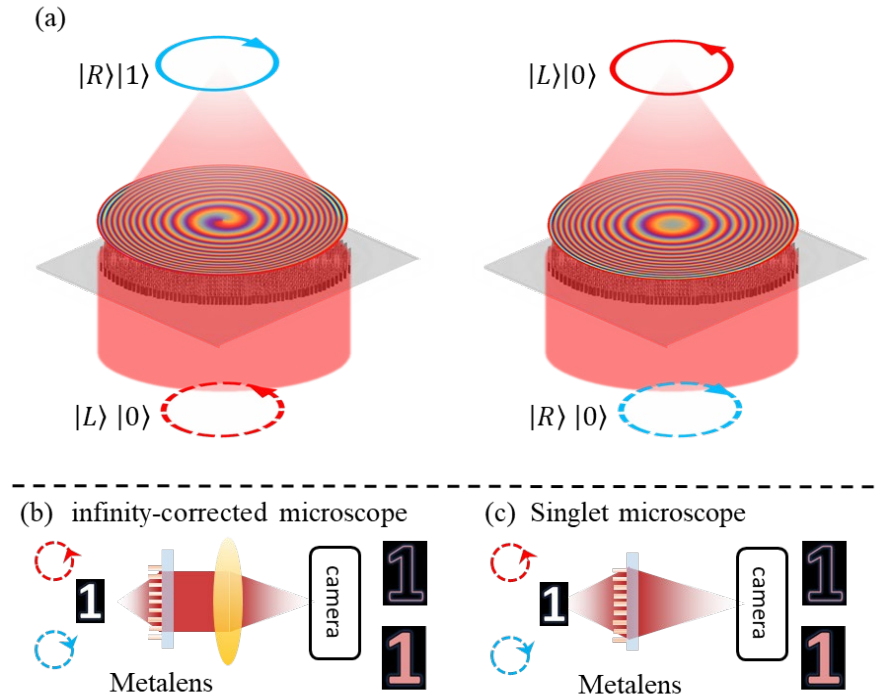

**Figure 1:** (a) Conceptual scheme of polarization-controlled focusing mode. Metalens-based infinity-corrected microscope (b) and singlet microscope (c) for polarization-controlled phase contrast imaging and bright field imaging.

$$\Psi_L = \frac{2\pi}{\lambda}\left(f - \sqrt{f^2 + r^2}\right) + \Phi \tag{1}$$

$$\Psi_R = \frac{2\pi}{\lambda}\left(f - \sqrt{f^2 + r^2}\right) \tag{2}$$

In Eqs. (1) and (2), $\varphi_x$ is the polarization-independent propagation phase of the nanofins array, $2\theta$ is the polarization-dependent PB phase. $\Psi_L$ is the phase profile imparted on the incident LCP light beam and $\Psi_R$ is the phase profile imparted on the incident RCP light beam. $f$ is the target focal length, and $(r, \Phi)$ is the polar coordinate of the nanofin overall metalens. As discussed in supporting material S1, the propagation phase $\varphi_x$ depends on the size ($W_x$ and $W_y$) of the nanofin [Fig. S1(a)], and $\theta$ is the rotation angle of the nanofin, Fig. S1b. A library including eight nanofins in Fig. S2d with height $H$ of 800nm and period $P$ of 650nm are designed to realize $2\pi$ phase coverage. The design details of the library can be seen in supplementary Fig S2.

Based on Eqs. (1) and (2), the target propagation phase profile $\varphi_x$ can be confirmed, Eq. S7 in supplementary material. Combining the nanofins library in Fig. S2d, the size of nanofin at each lattice $(r, \Phi)$ is determined. From Eqs. (1) and (2), the rotation angle $\theta$ of the nanofin at each lattice $(r, \Phi)$ is determined as well, Eq. (S8) in supplementary material. Repeating the above process at each lattice, the whole metalens is designed. A metalens was designed and fabricated with a diameter of 4mm, NA of 0.89 and a focal length $f$ of 1mm shown in Fig. 2a. The second metalens was designed and fabricated with a diameter of 4mm, NA of 0.25 and a focal length $f$ of 8mm, shown in Fig. 2d.

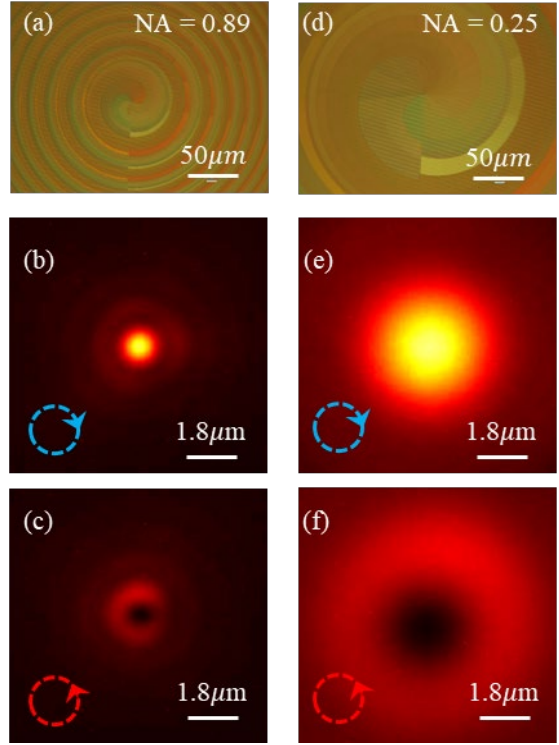

**Figure 2:** Characterization of metalens samples. The (a) – (c) are the optical image, Gaussian focal point, and donut-shaped focal point of the metalens sample 1 (NA = 0.89), respectively; (d) – (f) are the optical image, Gaussian focal point, and donut-shaped focal point of the metalens sample 2 (NA = 0.25), respectively.

The metalens samples were fabricated from the 800nm a-Si film which is deposited on a $SiO_2$ substrate via PECVD. The nanofins distribution is transferred to the a-Si film by using electron beam lithography, and the nanofins were etched by the reactive ion etcher. Figures 2a and 2d are the optical images of the two metalens samples. In the following, sample 1 refers to the metalens with a target NA of 0.89, and sample 2 refers to the metalens with a target NA of 0.25.

A customized microscope with a magnification of 100X and NA of 0.9 was used to characterize the polarization-controlled focusing ability of the two samples (Fig. S3 in supplementary material). The Gaussian focal point is first measured to evaluate the NA of two samples under the RCP beam illumination. The full width of half maximum (FWHM) of sample 1 is measured to be 1.1 $\mu m$, shown in Fig. 2b, which demonstrates that sample 1 has a NA of 0.89. The FWHM of sample 2's Gaussian focal point is 3.8$\mu m$ corresponding to a NA of 0.25, Fig. 2e. These results indicate that two diffraction-limited Gaussian focal points are obtained by illuminating samples 1 and 2 via an RCP light beam. As shown in Figs. 2c and 2f, the two Gaussian focal points were tuned to be two donut rings by changing the illumination light beam to LCP. Supporting video V1 shows the transition process of the focal field of metalens sample 1 by rotating a quarter waveplate (QWP), shown in Fig. S3 in supplementary material). The donut ring of sample 1 has a radius of 1.6 $\mu m$, and the donut ring of sample 2 has a radius of 5.7 $\mu m$, respectively.

To realize a large-magnification (58X) and a high-resolution (0.7$\lambda$) dual-mode imaging system, an infinity-corrected microscope in Fig. 1b was demonstrated by utilizing the metalens sample 1 as a high-NA (0.89) objective lens (Section 3.1). and, a second infinity-corrected dual-mode microscope with large FOV (600$\mu m$ × 800$\mu m$) and a diffraction-limited imaging resolution (0.61$\lambda$/NA) was demonstrated by utilizing the metalens sample 2 as the objective lens (Section 3.2). Finally, a singlet dual-mode microscope system in Fig. 1c was demonstrated via the metalens sample 2 (Section 3.3).

## 3. Compact dual-mode imaging experiments

### 3.1. High-resolution and large magnification infinity-corrected microscope

An infinity-corrected microscope in Fig. 1b was constructed by utilizing the metalens sample 1. The metalens serves as a high-NA (0.89) objective lens and polarization-controlled phase filter. A 1550nm LED light source (Thorlabs, M1550L4) is used for illuminating a USFA 1951 resolution target (Thorlabs, R1DS1N), and a cemented achromatic doublets lens (Thorlabs, AC254-060-C) with a focal length of 60mm is adopted as the tube lens. One polarizer and QWP are placed between the light source and the resolution target to manipulate the illumination light's polarization state. The image is captured by an InGaAs camera.

    Figures 3a and 3b show the bright-field image and edge-enhanced image of element 1 in group 7, respectively. Supplemented video V2 demonstrates the transition process of the imaging mode between bright-field and edge-enhanced phase contrast imaging mode by rotating the QWP. To quantitively demonstrate the transition of imaging mode, the intensity along the blue and red lines in Figs. 3a and 3b are depicted in Fig. 3c, which demonstrates the intensity peaks are tuned to the intensity valley with the imaging mode being tuned from bright-field to phase contrast imaging mode. According to the real dimension of element 1 in group 7, the magnification of our compact dual-mode microscope is calculated to be 58 times (i.e., 58X). The bright-field image and edge-enhanced image of the USAF element 6 in group 7 indicate that the microscope could effectively resolve a feature size around $1\mu m$ in both bright field and phase contrast imaging mode, which means the microscope has a diffraction-limited sub-wavelength resolution ($0.61\lambda/NA = 0.7\lambda = 1.1\mu m$) at the wavelength of 1550nm, shown in Fig S4a and S4b. However, limited by the sensor size of the InGaAs camera, this microscope can only provide us with a small FOV ($80\ \mu m \times 108\ \mu m$) in this configuration.

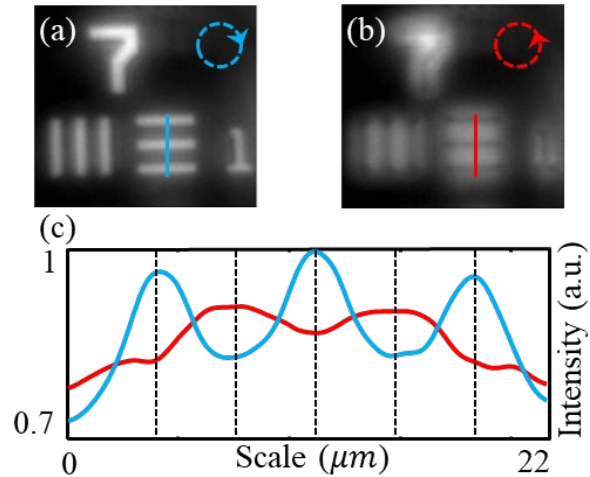

**Figure 3**: (a) and (b) show the BFI and EEI obtained via the 58X infinity-corrected microscope, and (c) demonstrates the intensity curve.

### 3.2. Diffraction-limited resolution and large FOV microscope for biological sample

We demonstrated the field of view of 600 μm× 800 μm, diffraction-limited image resolution, and biological dual mode imaging by using the metalens sample 2 with a focal length of 8 mm and NA of 0.25 in our custom build microscope. The imaging resolution and magnification of this microscope are firstly measured by capturing the dual-mode images of the group 7 in resolution target. Figures S4c and S4d respectively show the bright-field image and edge-enhanced image of the USAF group 7. Therefore, the imaging magnification and resolution of this microscope are calculated to be 7.8X and $3.78\mu m$ (i.e., diffraction limitation of the metalens sample 2, $0.61\lambda/NA=2.44\lambda$).

    A freshly peeled and unstained onion epidermal is prepared for verifying the dual-mode imaging ability of the microscope in biological samples. The cell wall (i.e., the edge of cells) becomes clear (Fig. 4b) from blurred (Fig. 4c) in whole FOV ($600\mu m \times 800\mu m$) with the illumination light is tuned to be LCP from RCP. The edge-enhancement process is demonstrated in supplemented video V3. The intensity along the blue and red lines in Figs. 4a and 4b are depicted in Fig. 4c to show the edge-enhancement effect quantitatively. In theory, the FOV can be further enlarged by adopting a camera with a larger imaging sensor.

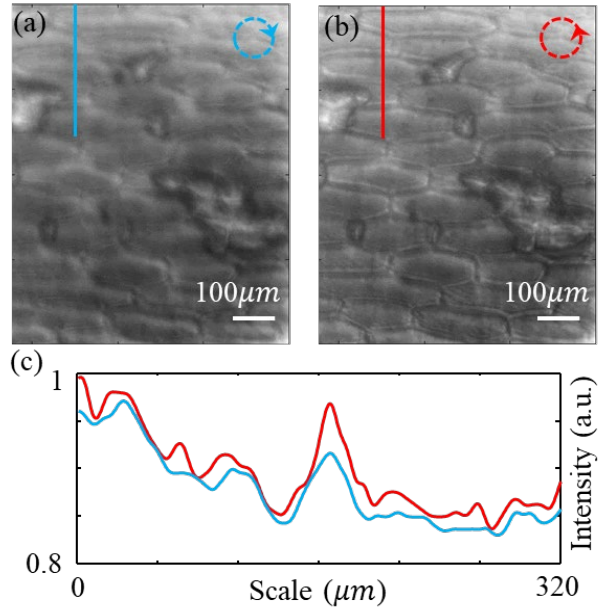

**Figure 4**: (a) and (b) show the bright-field image and edge-enhanced image obtained via the metalens-based 7.8X microscope, and (c) demonstrates the intensity curve.

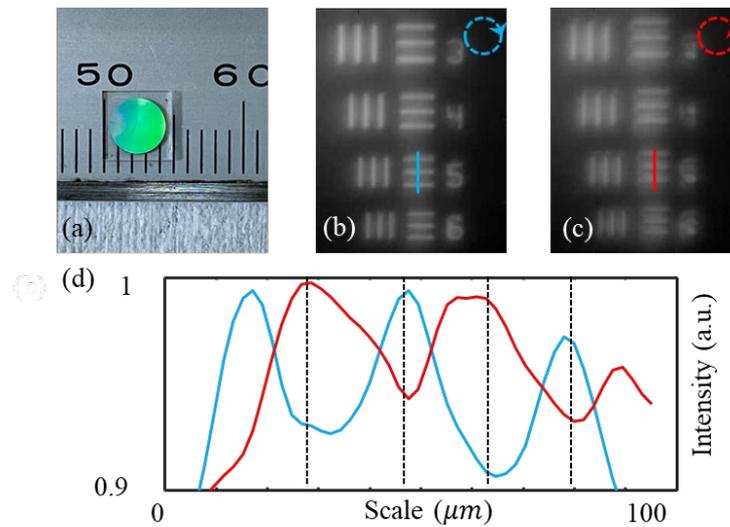

**Figure 5:** (a) metalens sample; bright-field image (b) and edge-enhanced image (c) obtained via the singlet dual-mode microscope, (d) is the intensity curve.

### 3.3. Singlet dual-mode microscope

While edge detection plays an important role in a lot of industrial applications, the edge-detection is mostly achieved by post-processing of a bright-field image via software algorithms. The processing speed of current edge-detection algorithms highly depends on the computer hardware and software. If we directly get the edge-enhanced images of the target objects based on an all-optics system, the edge detection would be improved to be light speed, which would boost the industrial application. In addition, the imaging system should be simplified as much as possible to improve the system's robustness and reduce the cost. In the following, a singlet dual-mode microscope is demonstrated to realize all-optical edge-detection at 1550 nm wavelength.

The metalens sample 2 is mounted 40mm in front of the InGaAs camera sensor (i.e., the imaging distance is 40mm) via a 3D-printed holder. The USFA 1951 resolution test target is placed 10mm in front of the metalens to get a clear image in the camera. Figures 5b and 5c show the bright-field image and edge-enhanced image of the elements 3-6 in group 5, respectively. According to the real dimension of element 5 in group 5, the magnification of this singlet dual-mode microscope is calculated to be 4.2X. The intensity curves in Fig. 5(d) demonstrate the intensity peaks are tuned to valleies with the imaging mode being tuned from bright-field to phase-contrast imageing mode. The tuning process is detailed and shown in supplemented video V4. Therefore, an all-optics edge-detection system with a simple hardware configuration is achieved in this research.

# 4. Conclusion

In summary, two compact dual-mode (i.e., phase-contrast and bright-field) microscope systems are demonstrated in the near-infrared spectrum based on silicon metalens. We first designed a bi-functional metalens which acts as a polarization-controlled phase plate and a high-NA objective lens. Two metalens samples with a diameter of 4mm are fabricated, and two samples respectively have a NA of 0.89 and 0.25. After characterizing the diffraction-limited focusing performance of two samples, the dual-mode infinity-corrected microscope was demonstrated by using the metalens samples as the objective lenses. The microscope can realize a large magnification of 58X and a high resolution of $0.7\lambda$ for the metalens sample 1, and a large field of view of $600\mu m \times 800\mu m$ is obtained by replacing the sample 1 with the sample 2. The edge-enhanced image of an unstained onion epidermal is observed by the microscope. Finally, a dual-mode singlet microscope is built based on the sample 2 for realizing all-optics edge-detection with light speed. Our experiment results illustrate that the dual-mode microscope systems have enormous potential for applications in biomedical diagnosis, biological and chemical research, and the industrial fields of machine vision and safety monitoring

**Acknowledgement**
PhD studentship from the Chinese Scholarship Council is acknowledged.
**Funding**
This work is supported by the UK funding agency EPSRC under grants EP/V000624/1, EP/X03495X/1, EP/T02643X/1 and the Royal Society RG\R2\232531
**Disclosure**
The authors declare no conflicts of interest.
**Data Availability**
The data that support the findings will be available in the University of Southampton's ePrints research repository following an embargo from the date of publication.